\documentclass[a4paper,11pt]{article}
\usepackage{pos}
\renewcommand{\logo}{\relax}
\graphicspath{{./}{../}}

\newcommand{\One}{1\kern-4.5pt1}

\title{Centre vortices in thermal lattice QCD with dynamical fermions}

\author[a]{Jackson Mickley}
\author*[b]{Chris Allton}
\author[c]{Ryan Bignell}
\author[a]{Derek Leinweber}

\affiliation[a]{Centre for the Subatomic Structure of Matter,
  Department of Physics, The University of Adelaide,
  South Australia 5005, Australia}
\affiliation[b]{Department of Physics, Swansea University,
       Singleton Park, Swansea SA2 8PP, UK}
\affiliation[c]{School of Mathematics \& Hamilton Mathematics Institute, Trinity College, Dublin, Ireland}
\emailAdd{c.allton@swansea.ac.uk}

\abstract{
Evidence for a second finite-temperature transition in Quantum Chromodynamics (QCD) is revealed through the study of centre vortex geometry and its temperature-dependent evolution. Using the anisotropic dynamical ensembles from the \textsc{Fastsum} collaboration, we conduct a detailed analysis at several temperatures above and below the established chiral transition.
Visualisations of the centre vortex structure in temporal and spatial lattice slices demonstrate that vortex percolation persists through the chiral transition and ceases at approximately twice the chiral transition temperature, \( T_c \). This indicates that confinement is maintained up to \( T \approx 2T_c \), suggesting a second transition associated with deconfinement. The loss of percolation, quantified by vortex cluster extent, provides a clear signal for this transition.
Statistical analysis of vortex density and cluster extent is presented which reveals consistent evidence of two transitions in QCD: the chiral transition at \( T_c \) and the deconfinement transition at \( T_d \approx 2T_c \). By analysing these, and other quantities, we obtain a precise determination of the deconfinement temperature, $T_d = 321(6) \,\text{MeV}$.
  }

\FullConference{The XVIth Quark Confinement and the Hadron Spectrum Conference (QCHSC24)\\
 19-24 August, 2024\\
 Cairns Convention Centre, Cairns, Queensland, Australia\\}

\begin{document}
\maketitle

\section{Introduction}

QCD's phase structure has been extensively studied using lattice simulations (see e.g.~\cite{Schmidt:2025ppy}). At low temperatures, QCD is confining, while asymptotic freedom suggests deconfinement at high temperatures. Early quenched simulations confirmed a first-order phase transition at $T_\text{quenched} \simeq 290$~MeV. However, with dynamical quarks, this transition becomes a crossover at the physical light quark masses, with $T_\text{c}^\text{phys} \simeq 158$ MeV, marking a chiral symmetry transition.
Conventionally, this crossover transition in the dynamical theory is considered to be the continuation of the quenched transition.

Centre vortices have long been used as a tool to understand the QCD confinement mechanism from an intuitive perspective~\cite{tHooft:1977nqb}.
It is therefore natural to investigate the 
evolution of centre vortex structure through the chiral transition and beyond.
In these proceedings, based on Ref. \cite{Mickley:2024vkm}, we use $2+1$ flavour anisotropic {\sc Fastsum} lattice ensembles, with light quarks corresponding to $m_\pi = 239(1)$ MeV, and temperatures ranging from 47 to 760 MeV.
The chiral crossover transition temperature for these ensembles
is $T_c = 167(3)$ MeV~\cite{Aarts:2020vyb,Aarts:2022krz}.
We gauge fix in a two-step procedure to the Maximal Centre Gauge and study the resultant centre vortex clusters.
By studying the vortex density and cluster extent we see clear evidence of confinement persisting through $T_c$ and a second QCD transition at $T\approx 2T_c$ which has deconfining properties. Indeed, there is some evidence suggesting that hadrons remain bound beyond $T_c$.
By studying several other quantities, as outlined in detail in Ref.~\cite{Mickley:2024vkm}, we determine a value of $T_d = 321(6)$ MeV.

\section{Two transition temperatures in the literature}
\label{sec:other}

Recent lattice QCD calculations with dynamical fermions have suggested hints of a second phase transition around $T/T_c \approx 2$, though its existence and location remains uncertain, with estimates varying between $200$ and $500$ MeV. While it seems established that $T_c$ is a chiral transition, there are some papers, summarised below, which postulate the presence of a second transition corresponding to a deconfinement temperature. This directly contradicts the widely-held picture that chiral symmetry restoration goes hand-in-hand with deconfinement.

Monopole condensation studies in $N_f=2+1$ QCD suggest a deconfinement phase, $T_{BEC}$, at $T \sim 275$ MeV~\cite{Cardinali:2021mfh}, aligning with pure gauge SU(2) and SU(3) results~\cite{DAlessandro:2010jdd,Bonati:2013bga}. Chiral-spin symmetry studies~\cite{Glozman:2022zpy, Glozman:2024ded, Chiu:2023hnm,Chiu:2024bqx} propose bound quark states above $T_c$, with distinct meson multiplets forming due to chromoelectric interactions. Defining the onset of these symmetries remains challenging due to their approximate nature.
Topological studies~\cite{Hanada:2023krw, Petreczky:2016vrs,Athenodorou:2022aay,Kotov:2021rah}, examining Polyakov loop observables and the topological charge, suggest a ``Partial Deconfinement" phase from approximately $0.9T_c$ to $1.9T_c$, with notable changes around $300$ MeV. Axion-motivated research links topological susceptibility changes to a dilute-instanton gas at $T \gtrsim 250$ MeV, consistent with quenched studies.
Scale invariance studies~\cite{Meng:2023nxf} further propose a new thermal QCD phase, observing spectral density shifts in the overlap Dirac operator from $200$ to $250$ MeV.

While this summary indicates a clear groundswell of work proposing a second transition temperature, none of these papers are able to provide a precise a value for this transition temperature.

\section{Lattice setup}
\label{sec:lattice}

\begin{table}[ht]
\begin{center}
 \begin{tabular}{|l|ccccccccccccc|}\hline
  $N_\tau$ & 128 &  64 &  56 &  48 &  40 &  36 &  32 &  28 & 24 &  20 & 16 &  12 & 8 \\ \hline
  $T\!$ (MeV)$\!\!$ & 47 & 95 & 109  & 127  & 152  & 169  & 190  & 217  &  253  & 304  & 380  & 507 & 760 \\ 
  $T/T_{c}$  & 0.28$\!\!$ & 0.57$\!\!$ & 0.65$\!\!$ & 0.76$\!\!$ & 0.91$\!\!$ & 1.01$\!\!$ & 1.14$\!\!$ & 1.30$\!\!$ & 1.52$\!\!$ & 1.82$\!\!$ & 2.28$\!\!$ & 3.03$\!\!$ & 4.55$\!\!$ \\ \hline
\end{tabular}
\caption{\label{tab:ensembles} Temporal extent $N_\tau$ and temperatures $T$ in MeV and in
  units of the chiral crossover temperature $T_c$ for the {\sc Fastsum} Generation 2L
  ensembles, used in this study \cite{Aarts:2022krz}.}
\end{center}
\end{table}

This study utilizes the ``Generation 2L'' thermal ensembles of the \textsc{Fastsum} collaboration \cite{Aarts:2022krz}, which employ \( 2+1 \) flavours of \( \mathcal{O}(a) \)-improved Wilson fermions on anisotropic lattices with a renormalized anisotropy of \( \xi = a_s/a_\tau = 3.453(6) \), where $a_{s(\tau)}$ is the spatial (temporal) lattice spacing, and
$a_s = 0.11208(31)$~fm~\cite{Wilson:2019wfr}.
The analysis is based on the ``Generation 2L'' ensembles with a pion mass of \( m_\pi = 239(1) \) MeV.
The strange quark mass is tuned to its physical value.
The lattice action follows that of the Hadron Spectrum Collaboration~\cite{Edwards:2008ja}, incorporating a Symanzik-improved gauge action with mean-field-improved coefficients and a mean-field-improved Wilson-clover fermion action with stout-smeared spatial links.
Full details of our lattice action can be found in Ref.~\cite{Edwards:2008ja}.
The ensembles were generated using \textsc{Openqcd-Fastsum}~\cite{glesaaen_jonas_rylund_2018_2216356}, an extension of \textsc{openQCD-1.6}.
A fixed-scale approach is used, where the cut-off $a_\tau$ is held fixed and the temperature is varied by changing \( N_\tau \) according to  
$T = \frac{1}{a_\tau N_\tau}$.
The study includes five ensembles below the chiral crossover transition temperature \( T_c = 167(3) \) MeV, one near \( T_c \), and seven above it. This broad temperature range is essential for the analysis.
Details of the temperature values and temporal extent, $N_\tau$, studied are in Table~\ref{tab:ensembles}. All of these temperatures used spatial volumes of $N_s= 32^3$ but some temperatures used $48^3$ as well, as described in Section \ref{sec:clusterextent}.
100 configurations were used for each temperature studied.

\section{Centre Vortices}
\label{sec:centre}

Centre vortices are topological objects in lattice QCD and they have been used for a long time to explain the mechanism of confinement~\cite{tHooft:1977nqb}.
Full details of our procedure to determine centre vortices are detailed in Ref.~\cite{Mickley:2024vkm}, but we summarise our approach here.

The key concept is the centre element of the SU(3) gauge group, $z\in \mathbb{Z}_3$, where
\[
    \mathbb{Z}_3 = \left\{ \exp\left(\frac{2\pi i}{3}\, m \right) \mathbb{I} \;\middle|\; m = -1,0,1 \right\} \,,
\]
are the cube roots of unity.
Centre elements commute with all elements of the SU(3) group, and this gives them a priviledged position.

We first fix the gauge to the ``Maximal Centre Gauge'' (MCG) by bringing each link as close as possible to one of the three centre elements. A two-step procedure is utilised, first gauge fixing isotropically to ensure spatial links are optimised and then anisotropically including the required square of the anisotropy ratio factor to weight time oriented links in the final gauge fixing~\cite{Mickley:2024vkm}.
We then define the centre element or "charge", $z_\mu(x)$, of each link as the closest centre element to the MCG-fixed link, $U_\mu^{MCG}(x)$, i.e. we project onto the centre element. We use the $z_\mu(x)$ to trivially factor the gauge-fixed link as follows,
\[
U_\mu^{MCG}(x) = z_\mu(x) \; R_\mu(x).
\]
Configurations comprising of just the ``residual'' $R_\mu(x)$ fields have been shown to have perturbative properties, whereas those comprised of just the centre charges, $z_\mu(x)$, can reproduce non-perturbative QCD features, such as a non-zero string tension~\cite{Biddle:2022zgw} and dynamical mass generation~\cite{Kamleh:2023gho}.

The centre "flux" through each plaquette is defined by $Z^P =  z_1\, z_2\, z_3^\ast\, z_4^\ast$ where the $z_i, i=1,\ldots 4$, are the centre charges of the four links around the plaquette $P$.
We are interested in plaquettes which are pierced by non-trivial fluxes, i.e. those where $Z^P = \exp(2\pi i \, m^P/3)$, with $m^P=\pm 1$.
The Gauss's Law equivalent for SU(3) constrains flux lines so that they form closed loops without ends.
These flux lines are termed ``centre vortices.'' A quark field taken along a closed path around one of these vortices will pick up an SU(3) phase equal to the centre charge of the vortex.
Because these centre vortices in the $z_{\mu}(x)$ fields can't be undone by a gauge transformation, they are topological objects, indelibly imprinted in the structure of the fields.
Note that due to the $Z(3)$ group nature of the centre charges, flux lines can branch.

From the physics point of view, it can be shown that ``percolating'' centre vortices reproduce the area law of the Wilson loop, meaning that they reproduce confinement~\cite{Engelhardt:1999fd}.
Strictly speaking, this is true only for percolations which pierce space-time plaquettes -- see later.

\section{Results}
\label{sec:results}

\subsection{Visualisations}
\label{sec:visualisations}

\begin{figure*}
    \includegraphics[width=0.48\linewidth]{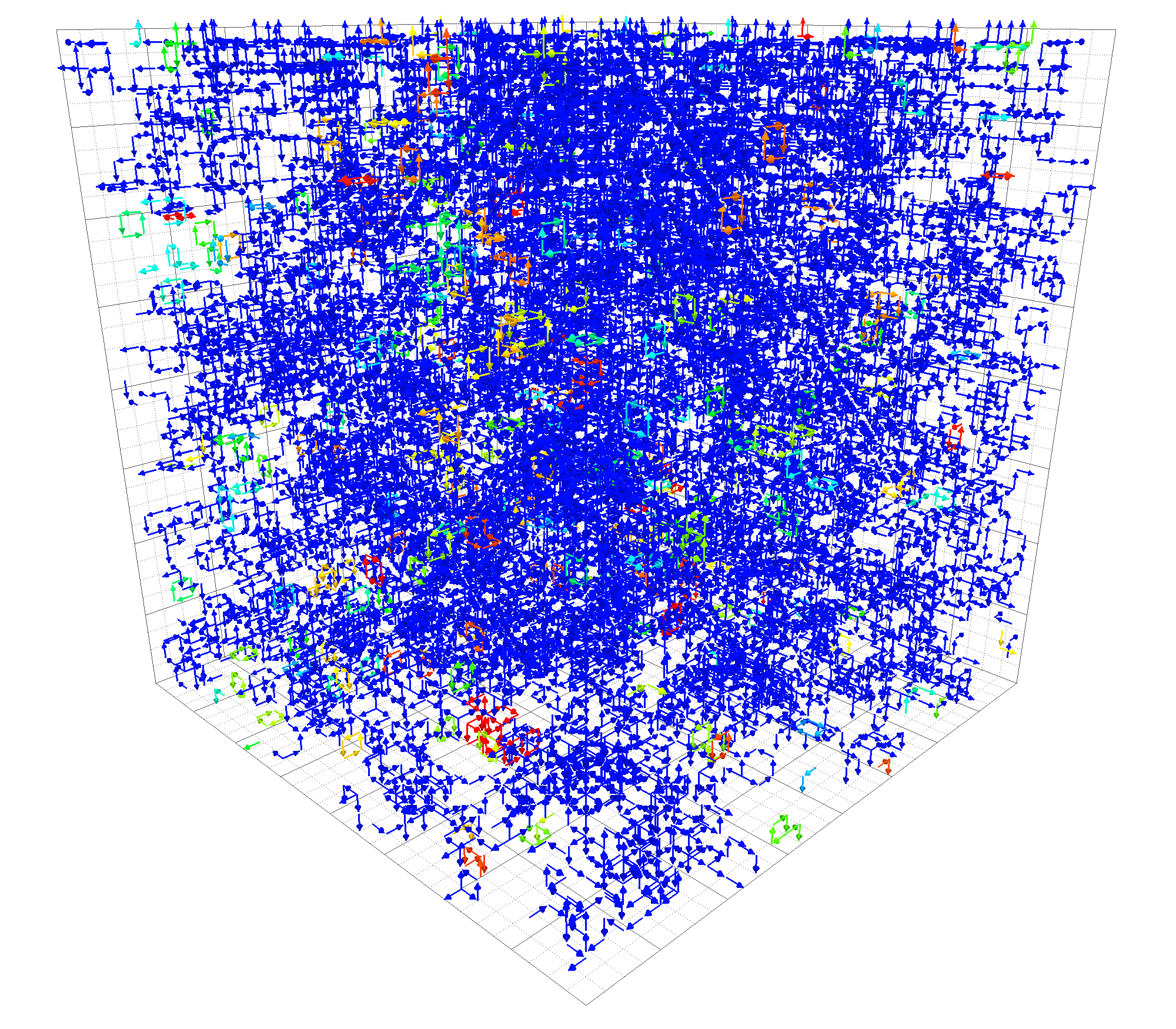}
    \includegraphics[width=0.48\linewidth]{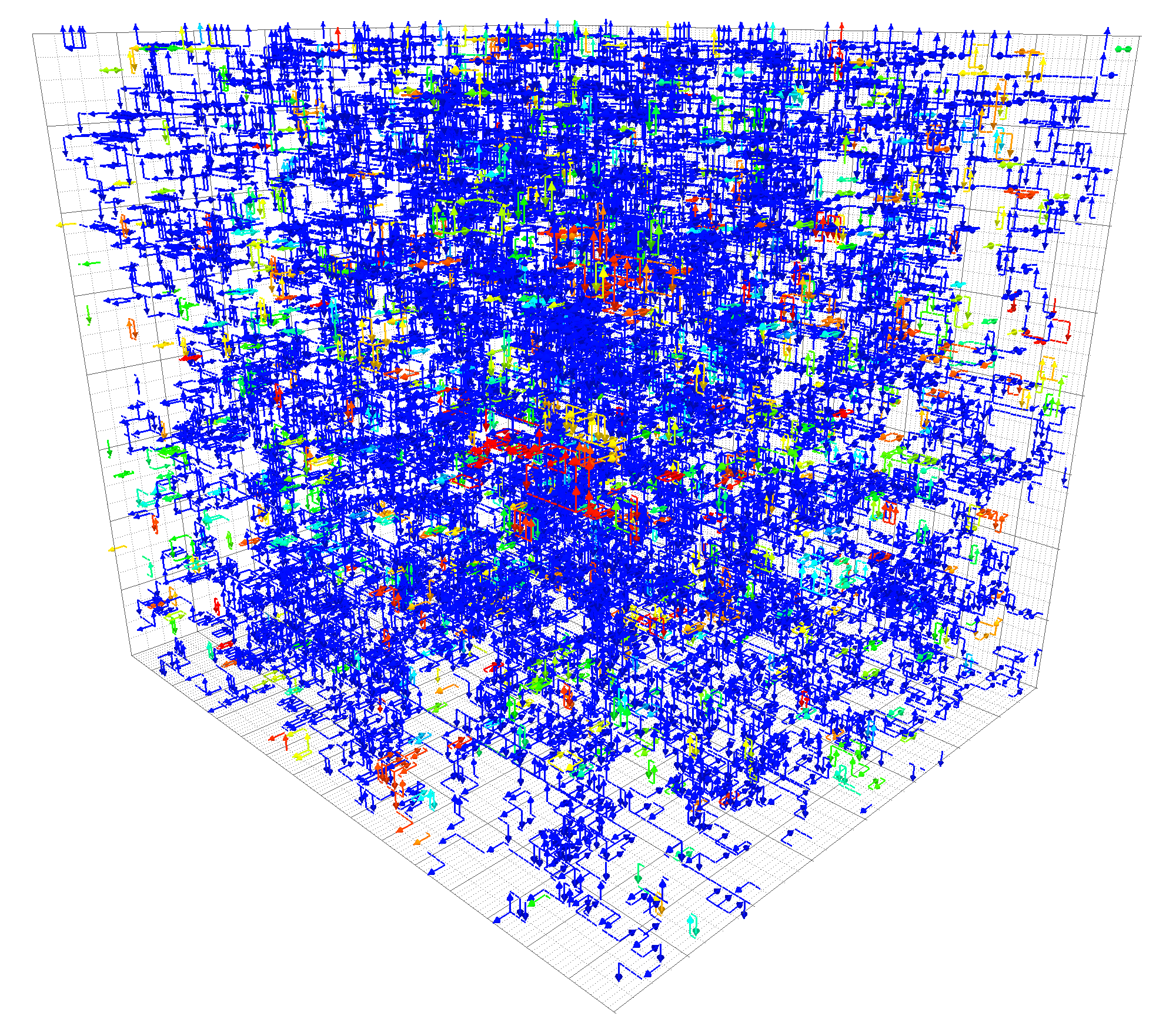}
    \vspace{-1em}
    \caption{\label{fig:Nt128vis} Centre vortex structure in temporal (\textbf{left}) and spatial (\textbf{right}) slices below the chiral transition temperature at $T/T_c \simeq 0.28$, corresponding to $N_\tau = 128$. In this and the following four figures, the compressed dimension in the right-hand figure is the temporal dimension. The significant reduction in jet length obscures the directional information of the arrowhead. The largest vortex cluster is coloured dark blue.}
\end{figure*}
\begin{figure*}
    \includegraphics[width=0.48\linewidth]{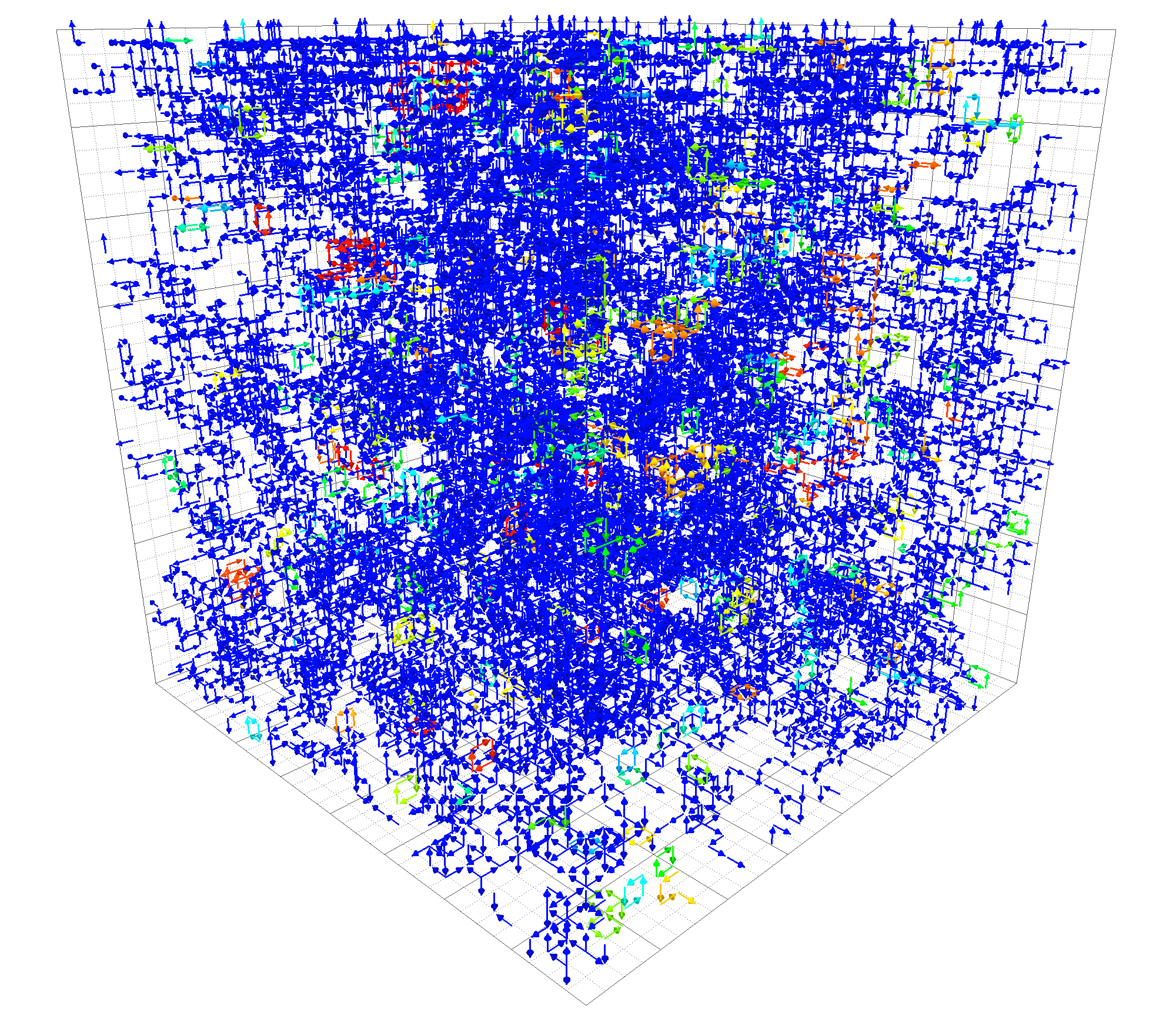}
    \includegraphics[width=0.48\linewidth]{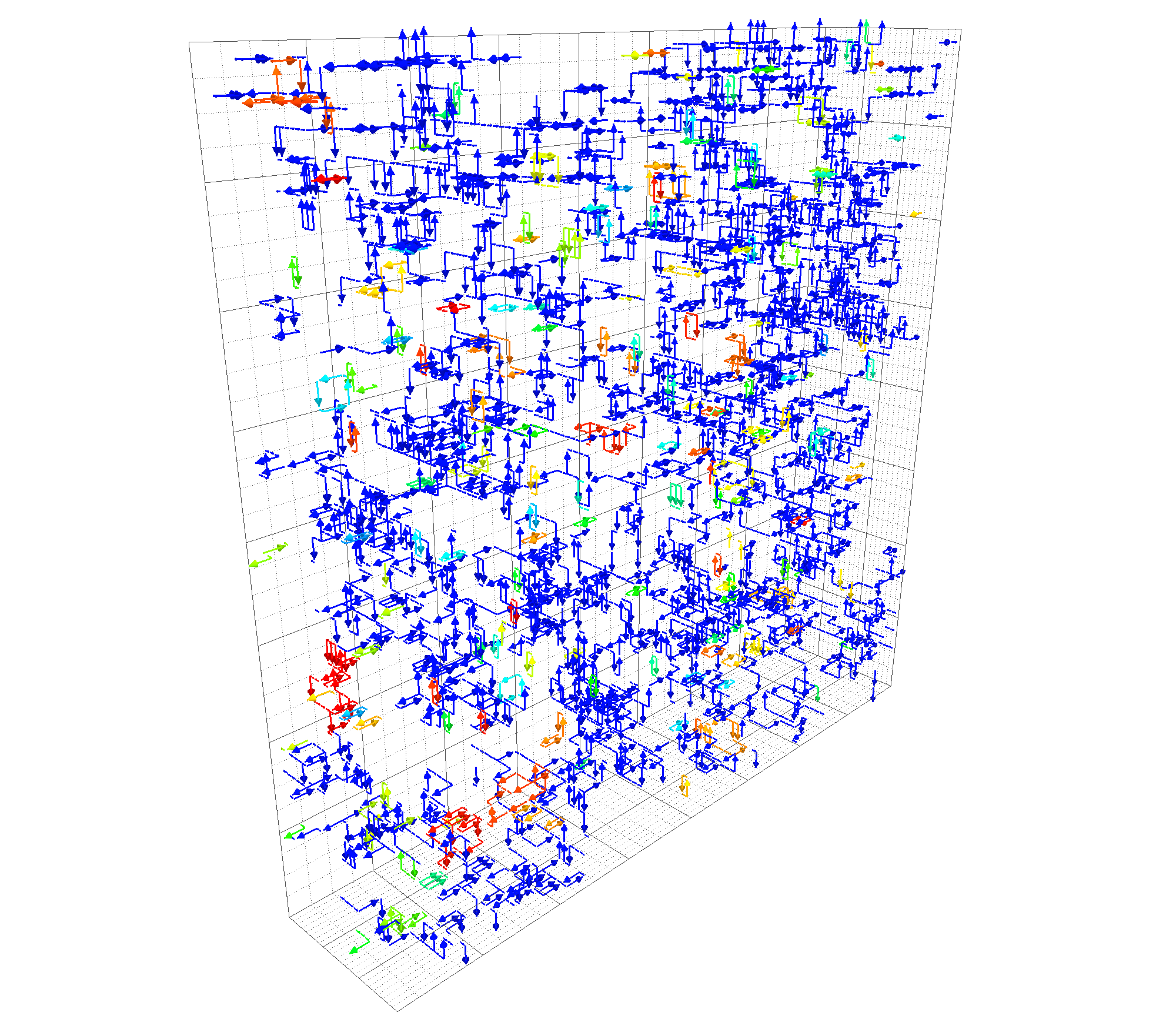}
    \vspace{-1em}
    \caption{\label{fig:Nt24vis} Centre vortex structure in temporal (\textbf{left}) and spatial (\textbf{right}) slices above the chiral transition at $T/T_c \simeq 1.52$, corresponding to $N_\tau = 24$.}
\end{figure*}

\begin{figure*}
    \includegraphics[width=0.48\linewidth]{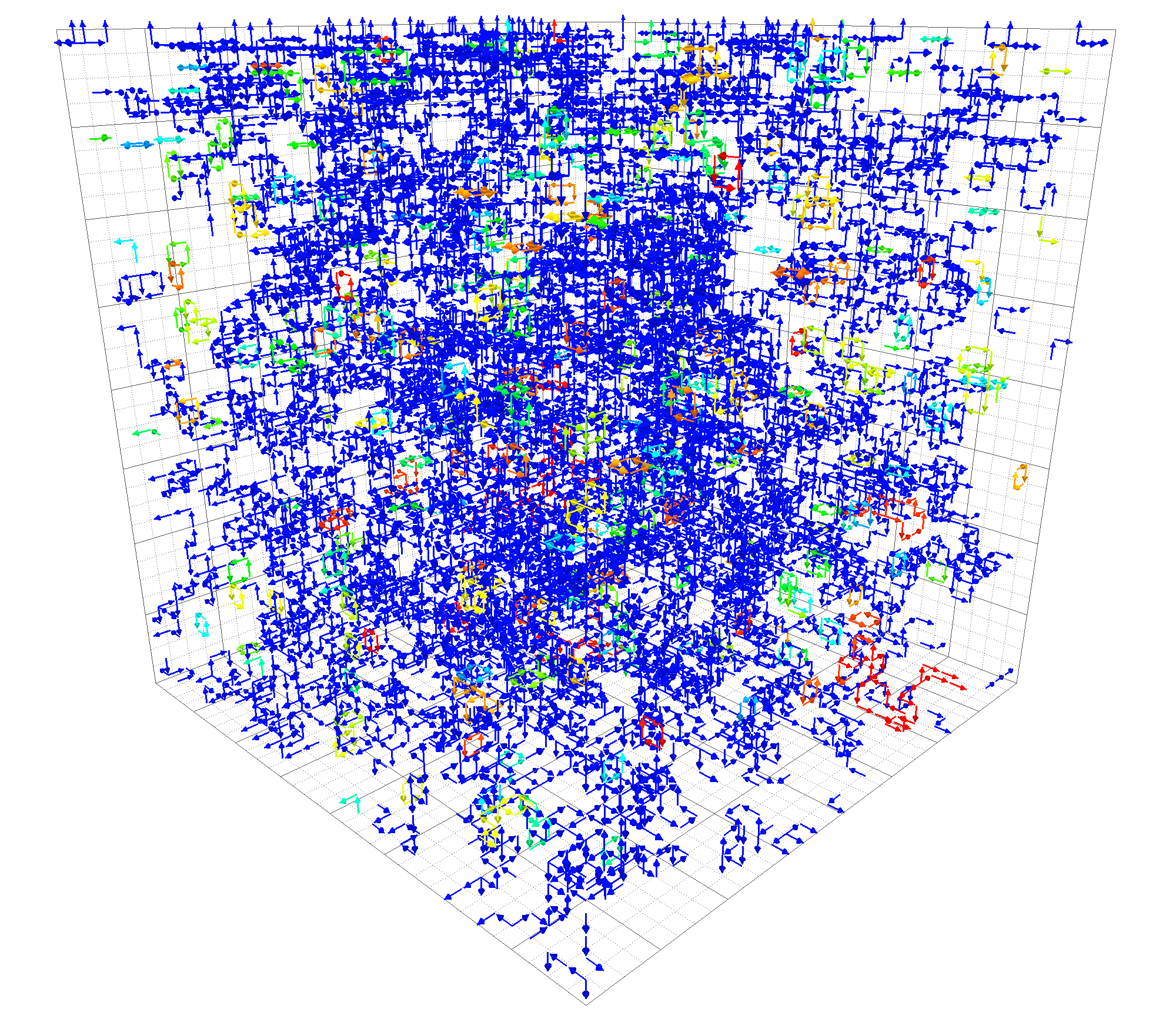}
    \includegraphics[width=0.48\linewidth]{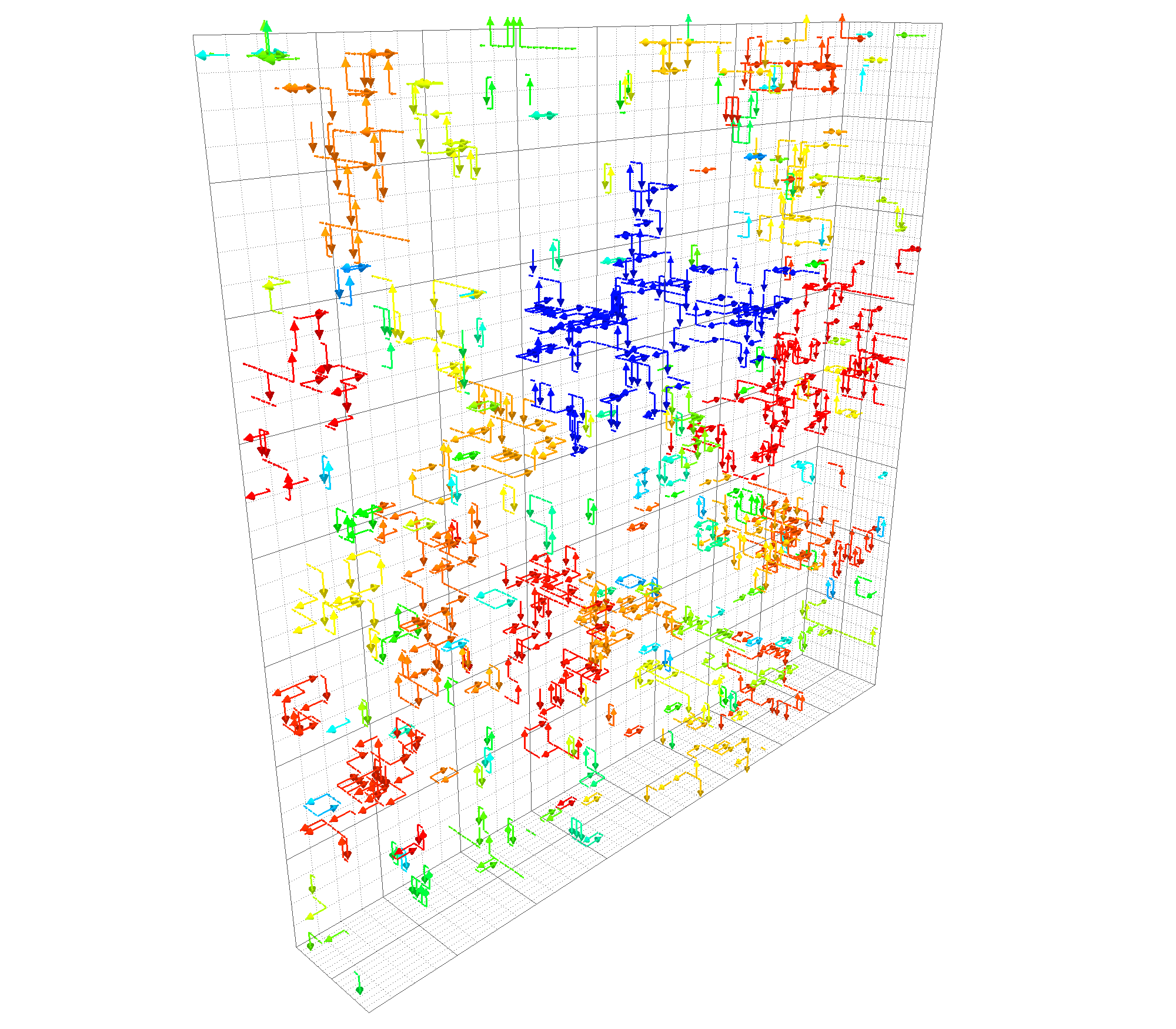}
    \vspace{-1em}
    \caption{\label{fig:Nt16vis} Centre vortex structure in temporal (\textbf{left}) and spatial (\textbf{right}) slices above the chiral transition at $T/T_c \simeq 2.28$, corresponding to $N_\tau = 16$.}
\end{figure*}

A useful way of understanding the effect of centre vortices on confinement is to visualise them at different temperatures.
Furthermore, comparisons of the dynamical quark visualisations with those from the quenched case can also be particularly useful to understand the subtleties of these theories.
However, since we are dealing with a four-dimensional space-time, visualisations are only practical if we hold one of the coordinates fixed and display the remaining three dimensions, e.g. by taking a slice at a fixed time or spatial coordinate.
In these visualisations, we represent a plaquette that is pierced by a vortex by an arrow going through the centre of the plaquette, i.e. on the dual lattice.
Since vortices form two-dimensional world-sheets (a vortex line in space traces out a two-dimensional sheet in space-time when propagated in time), a three-dimensional visualisation renders these vortex sheets as one-dimensional objects.
Because of Gauss's Law as discussed above, these objects form closed loops.

We use the techniques developed in Ref.~\cite{Biddle:2019gke} and display
vortices with $m^P = +1$ as arrows using a right-hand rule. $m^P = -1$ vortices are displayed as arrows in the opposite direction, meaning that the arrows represent the flow of $m^P = +1$ centre charge.

In Figures \ref{fig:Nt128vis}, \ref{fig:Nt24vis} and \ref{fig:Nt16vis} we display typical visualisations of temporal and spatial slices at temperatures $T/T_c= 0.28, 1.52$ and $2.28$ corresponding to $N_\tau = 128, 24$ and $16$.
These three temperatures lie in each of the three "phases" of QCD that will be discussed later.
A close inspection of the spatial slice visualisations shows that the temporal lattice spacing is correctly shown as being finer than the spatial one due to the anisotropy, $\xi = a_s/a_\tau \approx 3.5$.

In these visualisations, blue arrows are used for all vortices which are elements of the largest vortex cluster.
For the temporal sliced visualisations in Figs. \ref{fig:Nt128vis}, \ref{fig:Nt24vis} and \ref{fig:Nt16vis}, the largest clusters span the entire spatial volume, i.e. they percolate in $x,y,z$-space for all temperatures.

For the low temperature visualisation (Fig. \ref{fig:Nt128vis}) the vortex structures in the temporal and spatial slices are equivalent, indicating that there is percolation in the full four-dimensional space-time.
For the visualisations at the intermediate temperature, $T/T_c\simeq1.52$, (Fig. \ref{fig:Nt24vis}) the largest cluster clearly spans the three-dimensional volume displayed in both the temporal and spatial slices, indicating that percolation in the four dimensions still exists at this temperature, even though it is above $T_c$.
This contrasts starkly with results obtained in the pure-gauge theory where there is a sudden alignment of the vortex sheet with the temporal direction at the critical temperature, indicating the onset of deconfinement~\cite{Mickley:2024zyg}.
We therefore have qualitative evidence that confinement persists in full QCD above $T_c$.
At the highest temperature displayed, $T/T_c\simeq 2.28$, the structure of the vortices in the spatial slice changes.
Here, the largest cluster does not percolate, similar to the high temperature behaviour in the pure-gauge sector.
The system is populated by a large number of relatively small vortices.
If we consider the expectation value of a large space-time Wilson loop, $\langle W_{l\times t}\rangle$, all the small vortex clusters which are entirely inside the Wilson loop will not contribute to $\langle W_{l\times t}\rangle$.
It is only vortex clusters that are pierced by the Wilson loop that can contribute to $\langle W_{l\times t}\rangle$.
Since the vortex clusters are small at this temperature, these contributing vortex clusters will be positioned near the perimeter of the Wilson loop.
This naturally gives a perimeter law for $\langle W_{l\times t}\rangle$, using the identical argument that gives an area law for percolating clusters, as outlined in Ref.~\cite{Engelhardt:1999fd}.
This indicates that QCD is deconfined at this temperature.

We now turn to quantitative measures of the centre vortices.

\subsection{Centre vortex density}
\label{sec:vortexdensity}

\begin{figure}
    \includegraphics[width=0.48\linewidth]{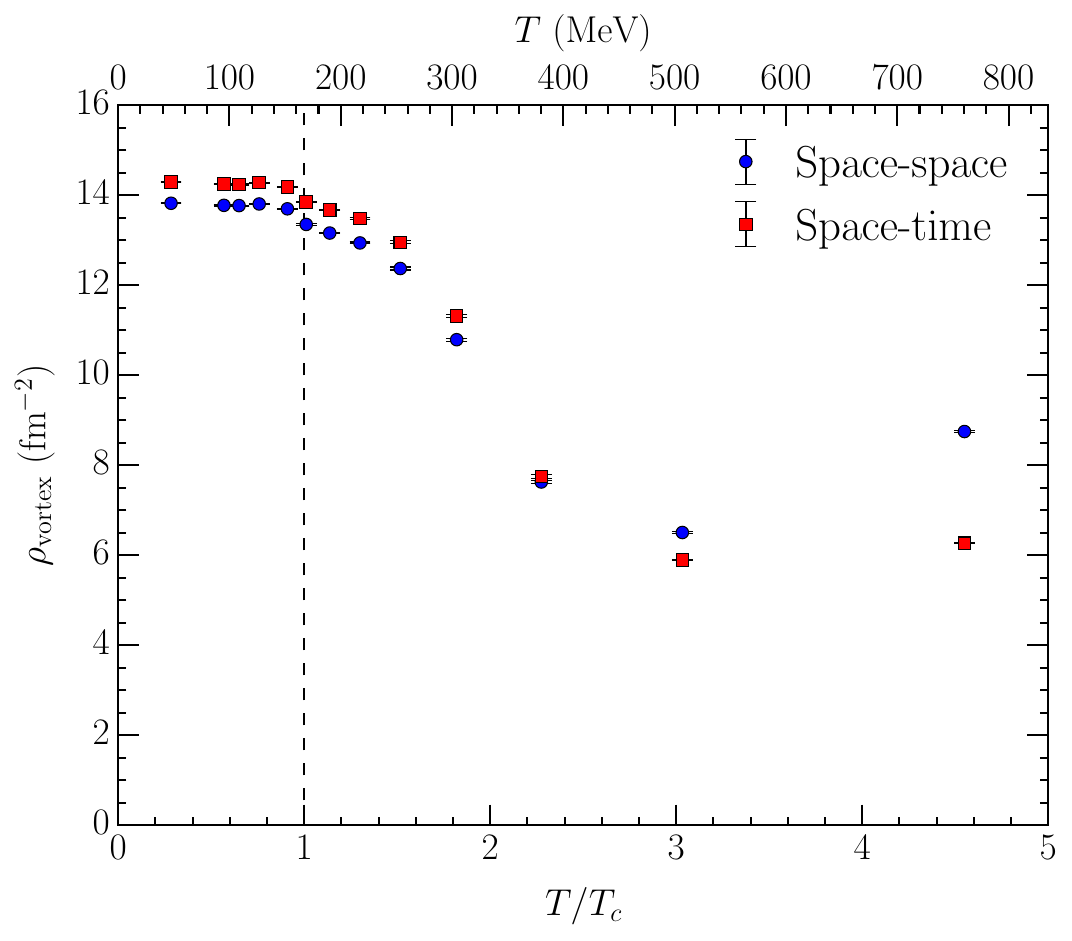}
    \includegraphics[width=0.5\linewidth]{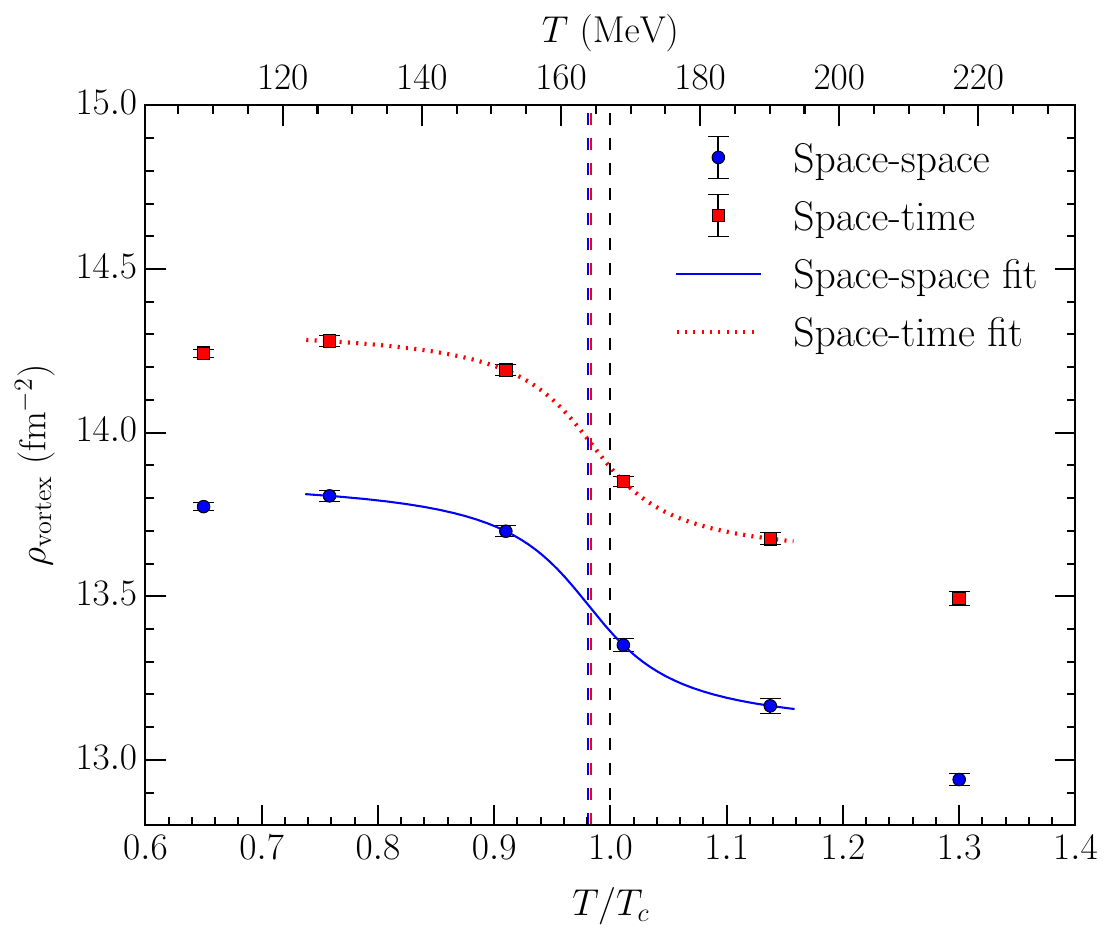}
    \caption{\label{fig:ss_st} (Left) The vortex density decomposed into space-space and space-time areas.
    The relative increase of space-space vortex density compared to space-time for $T/T_c \gtrsim 2$ can be attributed to the slight alignment of the vortex sheet with the temporal dimension which drives up the proportion of space-space plaquettes pierced.
    (Right) The space-space and space-time vortex densities zoomed in on $T_c$. There is visible crossover behaviour through the transition point, described by the sigmoid ansatz in Ref.~\cite{Mickley:2024vkm}. The resulting inflection points are strongly coincident with each other, and in reasonable agreement with $T_c$.}
\end{figure}

One of the easiest quantities to study is the centre vortex density, defined as the total number of plaquette piercings per unit physical area.
We calculate this density separately for space-space $(ss)$ and space-time $(st)$ oriented plaquettes, so the density is defined as,
$\rho_\mathrm{vortex}^{ss\,(st)} = \frac{N^\mathrm{pierced}_\mathrm{ss\,(st)}}{A_\mathrm{ss\,(st)}}$,
where $A_\mathrm{ss\,(st)} = a_s^2\; (a_sa_\tau)$ is the relevant plaquette area and $N^\mathrm{pierced}_\mathrm{ss\,(st)}$ is the total number of pierced $ss\,(st)$ oriented plaquettes.
Separating the $ss$ and $st$ vortex densities provides an important check that our approach has correctly included the anisotropy factors -- clearly in the infinite space-time volume limit, the physical density should be independent of how we discretise space and time.
For this reason, at the lowest temperatures where both the spatial and temporal lattice extent are sufficiently large, these densities should agree.

The vortex density results are displayed in Fig. \ref{fig:ss_st}.
Below $T_c$, the density remains fairly constant for both $ss$ and $st$ plaquettes.
Importantly they both agree within a few percent confirming that we have extracted the centre vortex content from our inherently anisotropic lattice in a physical, isotropic fashion.
It is interesting to speculate that this few percent discrepancy may be due to a small mis-tuning of the anisotropy, or at least due to the vortex density having a different lattice artefact to the quantities used to fix the quoted anisotopy, $\xi = 3.453(6)$.
Close inspection shows that the vortex density is sensitive to the chiral crossover at $T_c$ where there is a small drop.
The density then steadily decreases up to $T/T_c \approx 3$, before increasing again at higher temperatures.
We also note that the $ss$ and $st$ densities start to diverge at around $T\approx 2T_c$.

Figure \ref{fig:ss_st} (Right) shows a close-up  of the vortex density in the region of $T_c$.
Fitting this to a sigmoid function~\cite{Aarts:2020vyb,Burger:2018fvb,Aarts:2019hrg} reveals a point of inflection at
$T_1 = 0.981(8)(18)T_c$ (ss) and $T_1 = 0.983(8)(18)T_c$ (st)
where the first error is the statistical uncertainty from the inflection point fit, and the second is the contribution from the {\sc Fastsum} uncertainty on $T_c = 167(3)$~MeV.
Given that this first transition temperature $T_1$ equals $T_c$ within errors, these results confirm that the vortex density is able to resolve the chiral crossover transition.

We note from Fig. \ref{fig:ss_st} (Left) that the vortex density also shows signs of a crossover transition at around $T\approx 2T_c$.
Again fitting these results we find this second transition to be at
$T_2 = 1.913(6)(34)T_c$ (ss) and
$T_2 = 1.983(6)(36)T_c$ (st),
i.e. the two estimates are within $2\sigma$.
Since $T_2$ is aligns with the transition temperature between the intermediate and high temperature phases discussed in Section \ref{sec:visualisations}, we associate $T_2$ with the deconfinement temperature $T_d$.
In terms of vortex densities, this is marked by a difference appearing between the space-space and space-time densities due to the alignment of the vortex sheet with the temporal dimension.
Recall that percolating clusters which pierce space-time plaquettes are required for confinement.

\subsection{Cluster extent}
\label{sec:clusterextent}

\begin{figure}
    \centering
    \includegraphics[width=0.8\linewidth]{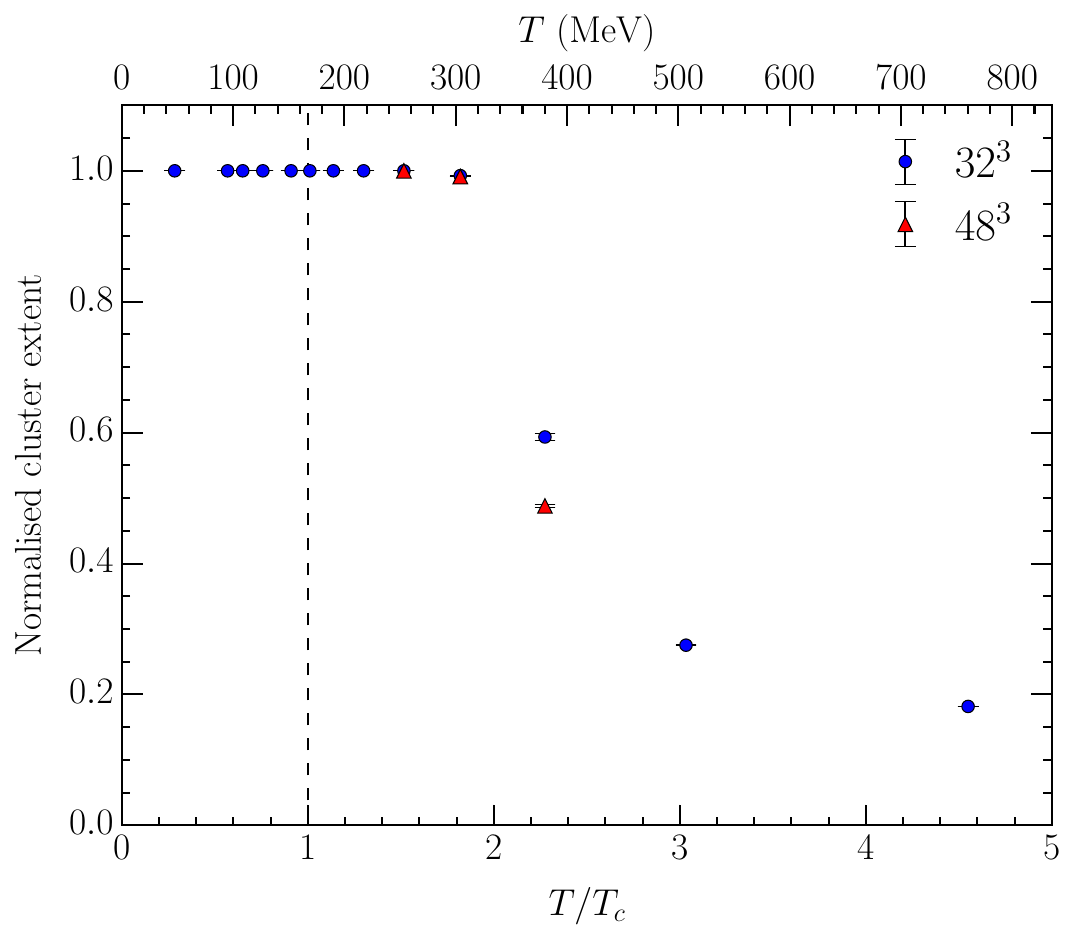}
    \caption{\label{fig:clusterextent} The normalised cluster extent for spatial slices. It attains a value of $\approx\! 1$ for all $T/T_c \lesssim 2$, indicating the vortex structure remains percolating until twice the chiral transition temperature. It thereafter rapidly decreases as percolation is lost.}
\end{figure}

A metric of vortex geometries which directly measures percolation is the normalised vortex extent.
As discussed in Section \ref{sec:visualisations} (see the temporal slices in Figs. \ref{fig:Nt128vis} -- \ref{fig:Nt16vis}) percolation in the spatial directions is observed for all temperatures.
We therefore only consider spatial slices which can access percolation, or lack of it, across both space and time dimensions.
The normalised cluster extent
is defined by first finding the greatest physical distance between any two points on the same vortex cluster, $d_\text{cluster}$, and then normalising this by the maximum distance it can take within our lattice volume, accounting for period boundary conditions.
For spatial slices, this maximum distance is $d_\text{max} = \frac{1}{2}\sqrt{ 2(a_s N_s)^2 + (a_\tau N_\tau)^2}$.
Finally, we average this quantity over all choices of spatial slices. 
Percolation implies normalised cluster extents of unity.

In Fig. \ref{fig:clusterextent} we see, unsurprisingly, that the normalised cluster extent is identically unity (with zero statistical errors) for $T<T_c$. 
Interestingly though, it remains unity above $T_c$ confirming percolation across space and time, and therefore confinement persists at these temperatures.
At $T/T_c=1.82$ we find a value marginally below unity, before dropping sharply away indicating a loss of percolation.

In order to check that our results are not contaminated by finite volume effects, we repeat our calculations using a spatial volume of $48^3$ at three temperatures centred on $1.82T_c$.
These results are also shown in Fig. \ref{fig:clusterextent}.
We see the the normalised cluster extent remains unity at $T/T_c=1.52$.
Interestingly, at $T/T_c=1.82$, both the $32^3$ and $48^3$ results agree.
We interpret this as evidence that there is a transition at around this temperature, confirming our other estimates of a second transition temperature.
At $T/T_c=2.28$, the normalised cluster extent decreases with volume.
Clearly $d_\text{max}$ increases with volume, so the more appropriate measure in this case is the unnormalised value, $d_\text{cluster}$ itself.
Our results show $d_\text{cluster}$ increases by around 23\%, which is much less that the increase of 50\% in physical extent of the lattice between these two volumes, see Ref.~\cite{Mickley:2024vkm} for details.

\section{Discussion and conclusion}
\label{sec:discussion}

\begin{figure}
    \centering
    \includegraphics[width=0.8\linewidth]{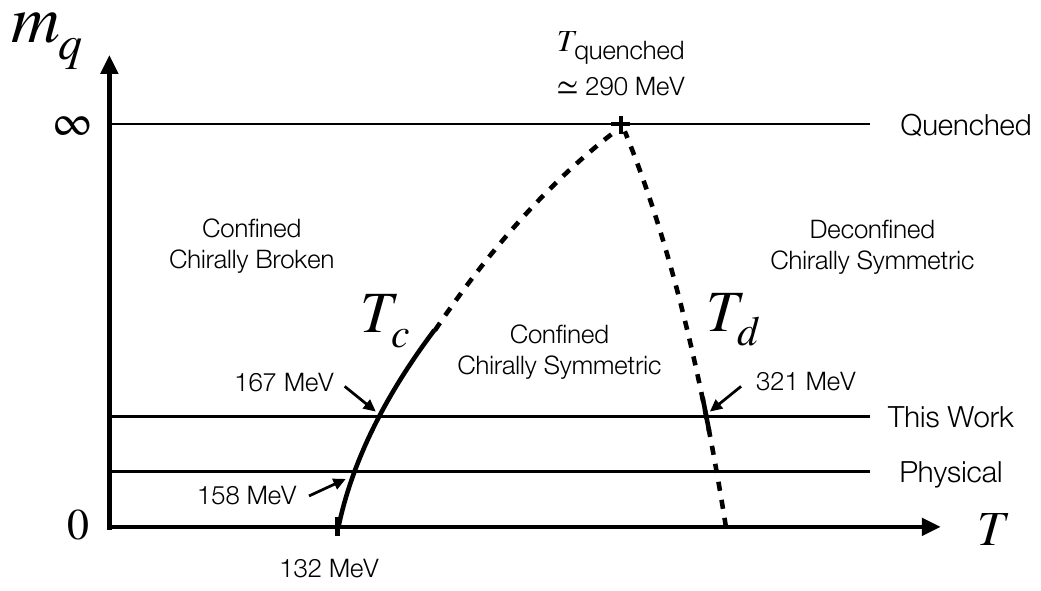}
    \caption{\label{fig:qcd-phase-diag}
     The proposed phase structure of QCD highlighting the two transition temperatures analysed in this work. Here, $m_q$ represents the masses of the light and strange dynamical quarks. The known dependence of the conventional chiral transition temperature, $T_c$, on $m_q$ is plotted as a solid curve, together with the proposed variation up to the quenched value, $T_\text{quenched}\simeq 290$~MeV~\cite{Borsanyi:2022xml}, as a dashed curve. The second transition temperature, $T_d=321(6)$~MeV, defined in this proceedings and in Ref.~\cite{Mickley:2024vkm} is displayed accompanied by its hypothesised variation with $m_q$ such that $T_d \rightarrow T_\text{quenched}$ in the quenched limit.
    These transition temperatures define three phases of QCD depending on the chiral symmetry and confinement properties.
    The values of $T_c=132(+6)(-3)$~MeV at the chiral point~\cite{HotQCD:2019xnw},
    $T_c \simeq 158$~MeV at the physical point~\cite{HotQCD:2018pds,Borsanyi:2020fev,Gavai:2024mcj}, and
    $T_c =167(3)$~MeV~\cite{Aarts:2022krz} at the quark masses considered here and in Ref.~\cite{Mickley:2024vkm} are shown.
}
\end{figure}

This proceedings has highlighted a few of the discoveries made in Ref.~\cite{Mickley:2024vkm} exploring centre vortex structure in dynamical, anisotropic QCD ensembles from the {\sc Fastsum} collaboration with 2+1 flavours of improved Wilson fermions with a pion mass of \( m_\pi = 239 \, \text{MeV} \).
The Maximal Centre Gauge and centre projection are used to obtain centre vortex configurations.
Visualisations of these vortices uncover interesting thermal behaviour, demonstrating that percolation in the space and time directions, and therefore confinement, persists above the chiral crossover transition, $T_c$, see Section \ref{sec:visualisations}.
These visualisations indicate deconfinement sets in at $T_d \approx 2T_c$.

In Section \ref{sec:vortexdensity}, the vortex density is studied.
Isotropic results are obtained for both space-space and space-time plaquettes to within a few percent.
Importantly the vortex densities uncover two transition temperatures,
the first in agreement with $T_c$, and the second at a temperature associated with deconfinement, $T_d \approx 2T_c$.

A key metric for percolation, the normalised cluster extent, is studied in Section \ref{sec:clusterextent}.
This confirms that percolation in space and time persists well above $T_c$ and ceases at around $2T_c$.
Careful studies are made to ensure that our conclusions are not contaminated by finite volume effects.

In our main paper,~\cite{Mickley:2024vkm}, we study other vortex quantities such as temporal correlation functions, branching point densities and probabilities, and the prevalence of secondary clusters.
In summary, all of these quantities confirm the picture outlined in these proceedings, i.e. that centre vortices uncover {\em two} transition temperatures in dynamical QCD:
\begin{itemize}
    \item The \textbf{chiral crossover transition} at $T_c = 164.2(1.4)$~MeV, consistent with previous results which marks a chiral transition, but with confinement persisting.
    \item The \textbf{deconfinement transition} at $T_d = 321(6)$~MeV, marked by the loss of centre vortex percolation in space and time, signalling the onset of deconfinement.
\end{itemize}
This contrasts with the quenched theory which exists in two phases, with the chiral and deconfinement transition co-existing.
The proposed QCD phase diagram is illustrated in Fig. \ref{fig:qcd-phase-diag} showing the merging of the transition temperatures $T_c$ and $T_d$ as the dynamical quark masses increase to the quenched limit.

Similar conclusions regarding an intermediate, confining phase of QCD have been proposed elsewhere (see Section \ref{sec:other}) albeit without accurate predictions for $T_d$.



\section*{Acknowledgments}

It is a pleasure to thank Dr.\ Waseem Kamleh for discussions regarding vortex identification on anisotropic lattices, and Jeff Greensite and Massimo D’Elia for other useful conversations.
This work was supported with supercomputing resources provided by the Phoenix HPC service at the University of Adelaide.
This research was undertaken with the assistance of resources and services from the National Computational Infrastructure (NCI), which is supported by the Australian Government.
We acknowledge EuroHPC Joint Undertaking for awarding the project EHPC-EXT-2023E01-010 access to LUMI-C, Finland.
This work used the DiRAC Data Intensive service (DIaL2 \& DIaL) at the University of Leicester, managed by the University of Leicester Research Computing Service on behalf of the STFC DiRAC HPC Facility (www.dirac.ac.uk).
This research was supported by the Australian Research Council through Grant No. DP210103706.
C.A. is grateful for support via STFC grant ST/X000648/1 and the award of a Southgate Fellowship from the University of Adelaide.
R.B. acknowledges support from a Science Foundation Ireland Frontiers for the Future Project award with grant number SFI-21/FFP-P/10186.
We are grateful to the Hadron Spectrum Collaboration for the use of their zero temperature ensemble.

\bibliography{bib}

\end{document}